\documentclass[10pt,letterpaper]{article}
\usepackage{amsmath}
\usepackage{amssymb}
\usepackage{graphicx}
\usepackage{bm}
\usepackage{opex3}
\usepackage{cite}

\renewcommand{\hbar}{h^{\hspace{-2.2mm}-}}

\begin{document}

\title{Hyperbolic metamaterial lens with hydrodynamic nonlocal response}

\author{Wei Yan,  N. Asger  Mortensen, and Martijn Wubs}

\address{Department of Photonics Engineering $\&$ Center for Nanostructured Graphene (CNG),  \\
Technical University of Denmark, DK-2800 Kongens Lyngby, Denmark}

\email{mwubs@fotonik.dtu.dk}

\begin{abstract}
We investigate the effects of  hydrodynamic nonlocal response in hyperbolic metamaterials (HMMs), focusing on the experimentally realizable parameter regime where unit cells are much smaller than an optical wavelength but much larger than the wavelengths of the longitudinal pressure waves of the free-electron plasma in the metal constituents. We derive the nonlocal corrections to the effective material parameters analytically, and illustrate the noticeable nonlocal effects on the dispersion curves numerically. As an application, we find that the focusing characteristics of a HMM lens in the local-response approximation and in the hydrodynamic Drude model can differ considerably. Interestingly, sometimes the nonlocal theory predicts significantly better focusing. Thus, to detect whether nonlocal response is at work in a hyperbolic metamaterial, we propose to measure the near-field distribution of a hyperbolic metamaterial lens.
\end{abstract}


\bibliographystyle{osajnl}

\section{Introduction}
Hyperbolic metamaterials (HMMs), also known as indefinite media, enjoy a great deal of attention owing to their unique hyperbolic dispersion relations~\cite{Smith:03,Smolyaninov:2011,Harish:2012,Jacob:2009,Poddubny:2011,Tumkur:2011,Noginov:2009,Belov:2006,Wood:2006,Li:2007,Salandrino:2006,Jacob:2006,Liu:2007,Yan:2009}, with associated high-wavenumber propagating waves without upper limit. This leads to numerous applications, such as enhanced-light interactions~\cite{Jacob:2009,Poddubny:2011,Tumkur:2011,Noginov:2009}, subwavelength imaging~\cite{Belov:2006,Wood:2006,Li:2007,Salandrino:2006,Jacob:2006,Liu:2007}, negative refraction~\cite{Smith:03}, and low-loss fiber cladding~\cite{Yan:2009}. The HMMs are usually artificially made by periodic dielectric-metal structures, such as a 1D dielectric-metal Bragg grating. To describe the optical properties of the HMMs, it is common to employ the local-response approximation (LRA), {\em i.e.} with every location ${\bf r}$ in the structure a certain value for the permittivity $\varepsilon({\bf r})$ is associated. In the LRA, the effective material parameters of HMMs  have been thoroughly studied~\cite{Elser:2007,Chebykin:2011,Chebykin:2012,Shen:2008,Shen:2007}.

Thanks to advances in nanofabrication, we witness a miniaturization of the feature size of metamaterials towards the deep nanoscale. The LRA becomes more inaccurate, since the nonlocal response of free electrons starts to play a role~\cite{Bloch:1933,Boardman:1982,Mochan:1987,David:2011,Abajo:2008,Raza:2011,Scholl:2012,Ciraci:2012b,Raza:2012b,Toscano:1,Toscano:2,FernandezDominguez:2012,Wei:2012}. It is known that the nonlocal response causes a blueshift of the surface plasmon (SP) resonance of the metallic particle~\cite{Raza:2011,Scholl:2012,Ciraci:2012b,Raza:2012b}, and limits the SP field enhancement that in the LRA sometimes diverges~\cite{Toscano:1,Toscano:2,FernandezDominguez:2012}.
We employ a simple generalization to the LRA, namely the hydrodynamic Drude model (HDM)~\cite{Bloch:1933,Boardman:1982}, which takes the nonlocal response into account. New in this model, as compared to the LRA, are longitudinal waves with sub-nanometer wavelengths, besides the usual transverse waves.

Incidentally, some HMMs have been found to  exhibit strong effective `nonlocal response', even in studies that employ the local-response approximation~\cite{Elser:2007,Shen:2007,Shen:2008,Chebykin:2011,Chebykin:2012}. What is meant here is that the HMMs, when considered as scatterers, cannot be described in the single-scattering Born approximation. This differs from the (material) nonlocal response of HMMs that we consider here and in Ref.~\cite{Wei:2012}, where already the scattering {\em potentials} associated with the metal constituents are nonlocal.

We recently showed that material nonlocal response plays an important role on HMMs in the limit of vanishing unit-cell size~\cite{Wei:2012}. In particular, the nonlocal response gives rise to a cutoff to the hyperbolic dispersion curve, and an associated finite but very large fundamental upper limit to the enhanced local optical density of states (LDOS), which in the LRA is known to diverge in the limit of vanishingly small unit cells. Now in realized HMMs, it is predominantly the finite size of the unit cell that keeps the LDOS finite. So it is an intriguing question, not explored yet as far as we know, whether nonlocal response can also have noticeable effects in state-of-the-art HMMs. This theoretical paper describes our search for observable nonlocal effects in HMMs with unit cells much smaller than an optical wavelength, but much larger than the wavelength of hydrodynamic longitudinal pressure waves.

\section{Dispersion relations of hyperbolic metamaterials}
As the HMM we consider a 1D subwavelength dielectric-metal Bragg grating, with a unit cell of thickness $d$, and thicknesses $a$ and $b$ of the dielectric and metal layers, respectively. The permittivity of the dielectric layer is $\epsilon_{\rm d}$.
The metal is described in the HDM as a free-electron plasma with~\cite{Bloch:1933,Boardman:1982,Raza:2011}
\begin{equation}
\epsilon_{\rm m}^{\rm T}(\omega) = 1 - \frac{{\omega_{\rm p}^2}}{{{\omega ^2} + i\omega\gamma}}, \qquad\qquad
\epsilon_{\rm m}^{\rm L}(k,\omega)= 1-\frac{\omega_{\rm p}^2}{\omega^2+i\omega\gamma-\beta^2k^2}, \label{epsTL}
\end{equation}
where $\epsilon_{\rm m}^{\rm T}$ is the Drude permittivity for the transverse electric fields as in the LRA, while the wavevector dependence of the permittivity $\epsilon_{\rm m}^{\rm L}$ of the longitudinal electric fields is responsible for the nonlocal response.

We apply the hydrodynamic generalization of the common transfer-matrix method for layered systems~\cite{Mochan:1987,Wei:2012}, and obtain the exact dispersion equation of the HMM
\begin{eqnarray}
\cos\theta_b&=&\Big\{ \cos {\theta _d}\Big[k_{{\scriptscriptstyle \bot}{\rm m}}^{\rm L}\cos {\theta _m}\sin {\theta _l} - \frac{{k_{\scriptscriptstyle \parallel}({w_d} - {w_m})}}{{{z_m}}}\sin {\theta _m}\cos {\theta _l}\Big] +\frac{{k_{\scriptscriptstyle \parallel}({w_d} - {w_m})}}{{{z_d}}}\sin {\theta _d}\nonumber\\
&\,&(1 - \cos {\theta _m}\cos {\theta _l})-\frac{1}{2}\Big[\frac{k_{\scriptscriptstyle \parallel}^2}{k_{{\scriptscriptstyle \bot}{\rm m}}^{\rm L}}\frac{{{}{{({w_d} - {w_m})}^2}}}{{{z_d}{z_m}}} + k_{{\scriptscriptstyle \bot}{\rm m}}^{\rm L}\left(\tfrac{{{z_d}}}{{{z_m}}} + \tfrac{{{z_m}}}{{{z_d}}}\right)\Big]\sin {\theta _m}\sin {\theta _l}\Big\} \nonumber\\
&\,&{\Big[k_{{\scriptscriptstyle \bot} {\rm m}}^{\rm L}\sin {\theta _l} - k_{\scriptscriptstyle \parallel}\frac{{({w_d} - {w_m})}}{{{z_m}}}\sin {\theta _m}\Big]^{ - 1}},\label{exact_dispersion}
\end{eqnarray}
where for convenience we introduced the dimensionless parameters
\begin{subequations}
\begin{eqnarray}
\theta_b&=&k_{\scriptscriptstyle \bot}d,\;\theta_d=k_{{\scriptscriptstyle \bot}\rm d}a,\;\theta_m=k_{{\scriptscriptstyle \bot}\rm m}^{\rm T}b,\;\theta_l=k_{{\scriptscriptstyle \bot}\rm m}^{\rm L}b,\\
z_{\rm d}&=&\frac{k_{{\scriptscriptstyle \bot}\rm d}}{k_0\epsilon_{\rm d}},\;w_{\rm d}=\frac{k_{\scriptscriptstyle\parallel}}{k_0},\;z_{\rm m}=\frac{k_{{\scriptscriptstyle \bot}\rm m}^{\rm T}}{k_0\epsilon_{\rm m}^{\rm T}},\;w_{\rm d}=\frac{k_{\scriptscriptstyle\parallel}}{k_0\epsilon_{\rm m}^{\rm T}},
\end{eqnarray}
\end{subequations}
and where $k_{\scriptscriptstyle \bot}$ represents the Bloch wavevector in the direction of the periodicity, $k_{\scriptscriptstyle\parallel}$  the wavevector along the layers, and $k_0=\omega/c$ the free-space wavevector. We also introduced the derived wavevectors $k_{{\scriptscriptstyle \bot}\rm d}=\sqrt{k_0^2\epsilon_{\rm d}-k_{\scriptscriptstyle\parallel}^2}$, ${k_{{\scriptscriptstyle \bot}\rm m}^{\rm T}}=\sqrt{k_0^2\epsilon_{\rm m}^{\rm T}-k_{\scriptscriptstyle\parallel}^2}$, and ${k_{{\scriptscriptstyle \bot}\rm m}^{\rm L}}=\sqrt{(\omega^2+i\gamma\omega-\omega_{\rm p}^2)/\beta^2-k_{||}^2}$.

We note that solving the hydrodynamic Drude model requires boundary conditions in addition to the usual Maxwell boundary conditions. The dispersion relation Eq.~(\ref{exact_dispersion}) was derived using the  continuity of the normal component of the free-electron current as the additional boundary condition. Details can be found in Refs.~\cite{Jewsbury:1981a,Raza:2011,Wei:2012}. The exact dispersion relation for nonlocal response Eq.~(\ref{exact_dispersion}) reduces to the exact dispersion relation in the LRA by putting the longitudinal permittivity $\epsilon_{\rm m}^{\rm L}(k,\omega)$ of Eq.~(\ref{epsTL}) equal to the transverse permittivity $\epsilon_{\rm m}^{\rm T}(\omega)$ of Eq.~(\ref{epsTL}), in other words by taking the nonlocal parameter $\beta$ to zero.

\section{Effective nonlocal material parameters}\label{Sec:effective}
As stated before, we consider sub-wavelength unit cells much smaller than optical wavelengths, but with metal layers much larger than the wavelengths of their longitudinal pressure waves. So we focus on the situation where the following two parameters are small,
\begin{equation}
k_0 d\ll 1,\,\quad
\frac{1}{|k_{\rm m}^{\rm L} b|} \ll 1 \label{condition},
\end{equation}
where $k_{\rm m}^{\rm L}=\sqrt{\omega^2+i\gamma\omega-\omega_{\rm p}^2}/\beta$ is the longitudinal wavevector.
Furthermore, we will only consider the frequency range $\omega<\omega_{\rm p}$, where $\epsilon_{\rm m}^{\rm T}<0$ and the dispersion curve of the HMM could be a hyperbola in the LRA.
We now make a first-order Taylor approximation in the small parameters of Eq.~(\ref{condition}) to the exact dispersion relation Eq.~(\ref{exact_dispersion}), and obtain the approximate dispersion relation
\begin{eqnarray}
k_{\scriptscriptstyle \bot}^2= k_0^2\epsilon_{\scriptscriptstyle \parallel}^{\rm loc}-k_{\scriptscriptstyle \parallel}^2\frac{\epsilon_{\scriptscriptstyle \parallel}^{\rm loc}}{\epsilon_{\scriptscriptstyle \bot }^{\rm loc}}-\left\{k_{\scriptscriptstyle \parallel}^2\frac{\epsilon_{\scriptscriptstyle \parallel}^{\rm loc}}{\epsilon_{\scriptscriptstyle \bot }^{\rm hdm}}+\Delta_{\rm la}\right\}.
\label{approx_dispersion}
\end{eqnarray}
The two terms in the brace are two leading correction terms that originate from the nonlocal response of the free electrons and from the finite size of the unit cell, respectively. Without them, {\em i.e.} when neglecting both the finiteness of the unit cells and nonlocal response, we have the well-known dispersion relation in the LRA
\begin{equation}
k_{\scriptscriptstyle \bot}^2=k_0^2\epsilon_{\scriptscriptstyle \parallel}^{\rm loc}-k_{\scriptscriptstyle \parallel}^2\frac{\epsilon_{\scriptscriptstyle \parallel}^{\rm loc}}{\epsilon_{\scriptscriptstyle \bot }^{\rm loc}},
\label{approx_dispersion2}
\end{equation}
where $\epsilon_{\scriptscriptstyle \parallel}^{\rm loc}$ and $\epsilon_{\scriptscriptstyle \bot}^{\rm loc}$ are the effective local permittivities~\cite{Wood:2006}
\begin{equation}
\epsilon_{\scriptscriptstyle \parallel}^{\rm loc} = f_{\rm d}\epsilon_{\rm d}+f_{\rm m}\epsilon_{\rm m},\qquad\epsilon_{\scriptscriptstyle \bot}^{\rm loc}=\frac{1}{f_{\rm d}\epsilon_{\rm d}^{-1}+f_{\rm m}\epsilon_{\rm m}^{-1}},\label{localmp}
\end{equation}
Obviously, when $\epsilon_{\scriptscriptstyle \parallel}^{\rm loc}\epsilon_{\scriptscriptstyle \bot}^{\rm loc}<0$, Eq.~(\ref{approx_dispersion2}) describes a hyperbolic dispersion curve.
However, we are now rather interested in nonlocal response effects in realistic HMMs, so we investigate the importance of the terms in the brace in Eq.~(\ref{approx_dispersion}), which are given in terms of
\begin{subequations}
\begin{eqnarray}
\epsilon _{\scriptscriptstyle \bot}^{\rm hdm}&=&\frac{{k_{\rm m}^{\rm L}d}}{{2i}}\frac{{\epsilon _{\rm m}^{\rm T}}}{{\epsilon_{\rm m}^{\rm T} - 1}}  \qquad\mbox{and}\label{ehdm}\\
{\Delta _{\rm la}}& =& \frac{1}{12}\epsilon _{\rm d}\epsilon _{\rm m}^{\rm T}{f_{\rm d}}{f_{\rm m}}{d^2}{\left( {k_0^2 - \frac{{k_{\scriptscriptstyle \parallel}^2}}{{{\epsilon _ {\scriptscriptstyle \bot}^{\rm loc} }}}} \right)^2}\nonumber\\
&\,&+\frac{1}{{12}}\epsilon _{\scriptscriptstyle \parallel} ^{\rm loc}{d^2}\left[ {2k_{{\scriptscriptstyle \bot}\rm  d}^2{{\left( {k_{ {\scriptscriptstyle \bot}\rm m}^{\rm T}} \right)}^2}{f_{\rm m}}{f_{\rm d}}\frac{{\epsilon _{\scriptscriptstyle \parallel} ^{\rm loc}}}{{{\epsilon _{\rm d}}\epsilon _{\rm m}^{\rm T}}} + {{\left( {k_{ {\scriptscriptstyle \bot} \rm d}} \right)}^4}\frac{{f_{\rm d}^3}}{{{\epsilon _{\rm d}}}} + k({{_{ {\scriptscriptstyle \bot} \rm m}^{\rm T}})^4}\frac{{f_{\rm m}^3}}{{\epsilon _{\rm m}^{\rm T}}}} \right]. 
\end{eqnarray}
\end{subequations}
From the expression~(\ref{ehdm}) for $\epsilon_{\scriptscriptstyle \bot}^{\rm hdm}$, it follows that the local-response limit is found in the limit ${k_{\rm m}^{\rm L}d} \to \infty$ rather than by taking the size $d$ of the unit cell to zero, see also Ref.~\cite{Wei:2012} on this point.
In the following we will keep the first ({\em i.e.} the nonlocal) correction term, while neglecting the other correction term ${\Delta _{\rm la}}$. The sole justification for doing so is that the second term turns out to be negligible in comparison to the first one for the HMMs that we consider, as we illustrate numerically below.
So, as an improved approximation to Eq.~(\ref{approx_dispersion2}), we keep the correction term concerning $\epsilon_{\scriptscriptstyle \bot }^{\rm hdm}$, and then obtain the approximate nonlocal dispersion relation that is central to our present study,
\begin{eqnarray}
k_{\scriptscriptstyle \bot}^2= k_0^2\epsilon_{\scriptscriptstyle \parallel}^{\rm loc}-k_{\scriptscriptstyle \parallel}^2\frac{\epsilon_{\scriptscriptstyle \parallel}^{\rm loc}}{\epsilon_{\scriptscriptstyle \bot }^{\rm nloc}},
\label{approx_dispersion3}
\end{eqnarray}
in terms of the new effective permittivity in the direction of the periodicity corrected by the nonlocal response
\begin{eqnarray}
\frac{1}{\epsilon_{\scriptscriptstyle \bot }^{\rm nloc}}=\frac{1}{\epsilon_{\scriptscriptstyle \bot }^{\rm loc}}+\frac{1}{\epsilon_{\scriptscriptstyle \bot }^{\rm hdm}}.
\label{nlocapmp}
\end{eqnarray}
The other tensor component $\epsilon_{\scriptscriptstyle \parallel}^{\rm loc}$ of the dielectric tensor  has no such nonlocal correction.

Since $1/|k_{\rm m}^{\rm L} d|\ll1$, we usually have $\epsilon_{\scriptscriptstyle \bot }^{\rm loc}\ll\epsilon_{\scriptscriptstyle \bot }^{\rm hdm}$, and then $\epsilon_{\scriptscriptstyle \bot }^{\rm nloc}\approx\epsilon_{\scriptscriptstyle \bot }^{\rm loc}$ is a good approximation. However, for large $\epsilon_{\scriptscriptstyle \bot}^{\rm loc}$ and especially in the extreme case that $\epsilon_{\scriptscriptstyle \bot}^{\rm loc}\to \infty$, it may occur that $\epsilon_{\scriptscriptstyle \bot }^{\rm loc}$ becomes much larger than $\epsilon_{\scriptscriptstyle \bot }^{\rm hdm}$, so that  $\epsilon_{\scriptscriptstyle \bot }^{\rm nloc}\approx\epsilon_{\scriptscriptstyle \bot }^{\rm hdm}$ by virtue of Eq.~(\ref{nlocapmp}). So in this case, the nonlocal response has the important role of replacing the infinite
$\epsilon_{\scriptscriptstyle \bot}^{\rm loc}$ by the finite $\epsilon_{\scriptscriptstyle \bot}^{\rm hdm}$ in the dispersion relation.
But when does it occur that $\epsilon_{\scriptscriptstyle \bot}^{\rm loc}\to \infty$? This immediately follows from Eq.~(\ref{localmp}), and occurs when $f_{\rm d}\epsilon_{\rm}^{\rm T}+f_{\rm m}\epsilon_{\rm d}$ vanishes. In particular, neglecting the Drude damping $\gamma$ of the metal,
the frequency $\omega_{\rm res}^{\rm loc}$ for which  $\epsilon_{\scriptscriptstyle \bot}^{\rm loc}\to \infty$ is given by
\begin{equation}
\omega_{\rm res}^{\rm loc}=\omega_{\rm p}\sqrt{\frac{f_{\rm d}}{f_{\rm d}+f_{\rm m}\epsilon_{\rm d}}}.
\label{res1}
\end{equation}
Let us also consider another extreme case, namely $\epsilon_{\scriptscriptstyle \bot}^{\rm nloc}\to \infty$. This is achieved  when $\epsilon_{\scriptscriptstyle \bot}^{\rm loc}=-\epsilon_{\scriptscriptstyle \bot}^{\rm hdm}$, which happens at the frequency
\begin{equation}
\omega_{\rm res}^{\rm nloc}\approx\omega_{\rm res}^{\rm loc}\left(1+\frac{\epsilon_{\rm d}}{k_{\rm m}^{\rm L} a}\right).
\label{res2}
\end{equation}
At this frequency $\omega_{\rm res}^{\rm nloc}$, the nonlocal response changes the finite $\epsilon_{\scriptscriptstyle \bot}^{\rm loc}$ into the infinite $\epsilon_{\scriptscriptstyle \bot}^{\rm nloc}$, again a strong nonlocal effect.

From the above analysis, it follows that nonlocal response  is important for HMMs near the frequencies $\omega_{\rm res}^{\rm loc}$ and $\omega_{\rm res}^{\rm nloc}$, as also evidenced numerically in the next section. The larger the difference between $\omega_{\rm res}^{\rm loc}$ and $\omega_{\rm res}^{\rm nloc}$, the broader the frequency range with noticeable nonlocal effects. As indicated by Eqs.~(\ref{res1}) and~(\ref{res2}), the nonlocal resonance frequency $\omega_{\rm res}^{\rm nloc}$ is blueshifted with respect to the local one $\omega_{\rm res}^{\rm loc}$. The relative blueshift $(\omega_{\rm res}^{\rm nloc} -\omega_{\rm res}^{\rm loc})/\omega_{\rm res}^{\rm loc}$ is proportional to $\epsilon_{\rm d}$, and inversely proportional to the dimensionless parameter $k_{\rm m}^{\rm L}a $. The latter dependence  is  rather surprising, since $k_{\rm m}^{\rm L}$ is the longitudinal wavevector of the free electrons in the metal layers, whereas $a$ is the thickness of the dielectric layer! We checked, also numerically, that our first-order Taylor expansion of the dispersion relation in the small parameter $(k_{\rm m}^{\rm L}b)^{-1}$ indeed gives a relative blueshift $\epsilon_{\rm d}/(k_{\rm m}^{\rm L} a)$, independent of the thickness $b$ of the metal layer.

\section{Effects of nonlocal response on the dispersion curve: numerical analysis}
To numerically illustrate the effects of the nonlocal response on the HMM, we choose a specific example of the HMM with $a=6\rm nm$, $b=3\rm nm$, and $\epsilon_{\rm d}=10$. We choose the metal to be  Au, and describe it by only its free-electron response, with parameters $\hbar \omega _{\rm {p}}= 8.812 \rm{eV}$, $\hbar {\gamma} = 0.0752 \rm{eV}$, and $v_{\rm F}=1.39\times10^6\rm{m/s}$.
\begin{figure}[hb]
\centering
\includegraphics[width=0.6\textwidth]{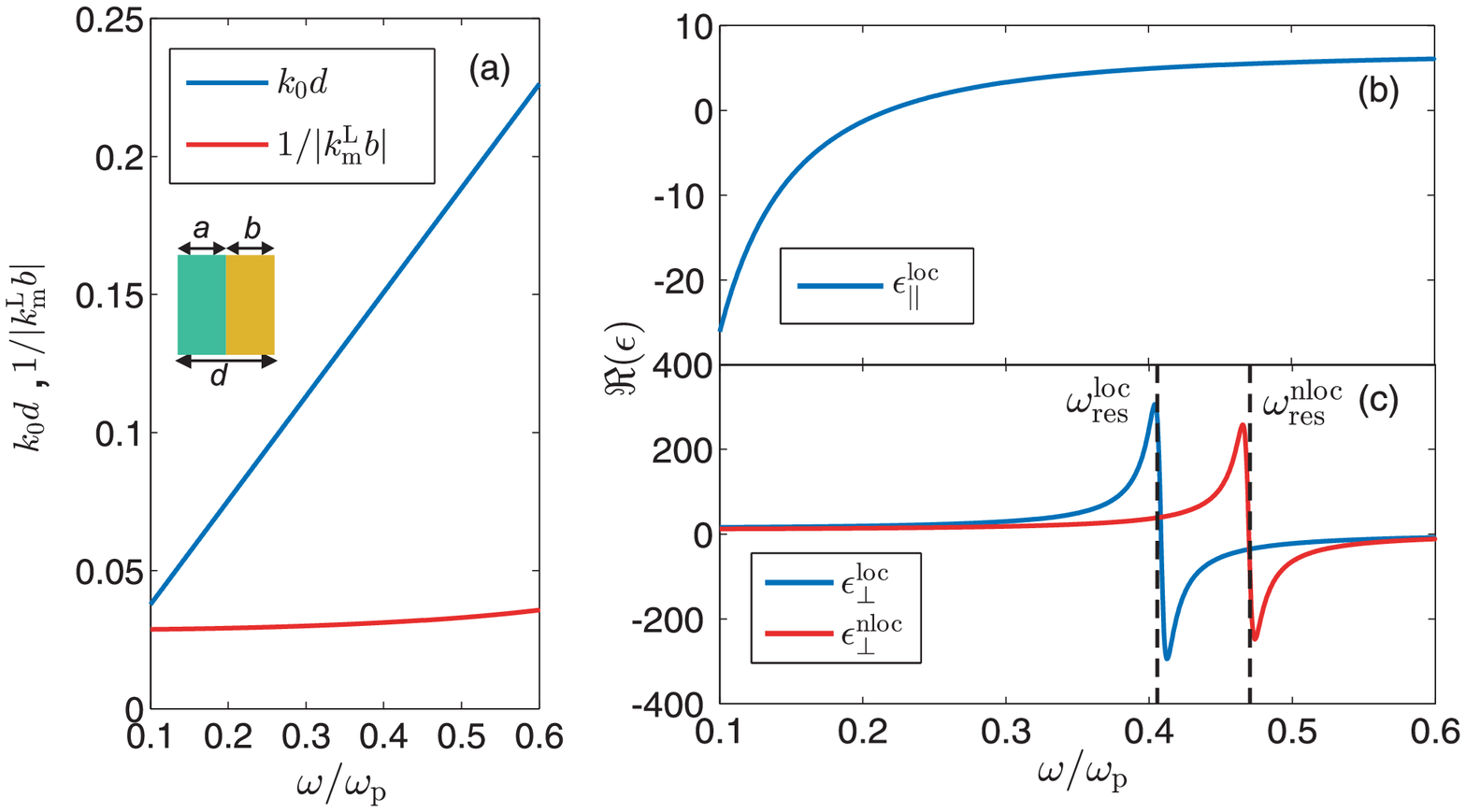}
\caption
{(a) $k_0d$ and $1/|k_{\rm m}^{\rm L}b|$, (b) real part of $\epsilon_{||}^{\rm loc}$, and (c) real parts of $\epsilon_{\scriptscriptstyle \bot}^{\rm loc}$ and $\epsilon_{\scriptscriptstyle \bot}^{\rm nloc}$, of the HMM. The unit cell of the HMM shown in the inset of (a) has $a=6\rm nm$, $b=3\rm nm$, $\epsilon_{\rm d}=10$, and the yellow metal layer is Au.}
\label{fig2}
\end{figure}

Fig.~\ref{fig2}(a) shows the value of $k_0 d$ and $1/|k_{\rm m}^{\rm L} b|$. Clearly, in the frequency range $0.1\omega_{\rm p}<\omega<0.6\omega_{\rm p}$, both $k_0 d$ and $1/|k_{\rm m}^{\rm L} b|$ satisfy Eq.~(\ref{condition}). This indicates that we are in the regime where the analytical results of the above section are valid. Fig.~\ref{fig2}(b) displays the real part of $\epsilon_{||}^{\rm loc} = \epsilon_{||}^{\rm nloc}$, and Fig.~\ref{fig2}(c) the real parts  $\epsilon_{\scriptscriptstyle \bot}^{\rm loc}$ and $\epsilon_{\scriptscriptstyle \bot}^{\rm nloc}$. The two dashed vertical lines in Fig.~\ref{fig2}(c) mark the positions of $\omega_{\rm res}^{\rm loc}$ and $\omega_{\rm res}^{\rm nloc}$ as calculated with the approximate but accurate Eqs.~(\ref{res1}) and (\ref{res2}), respectively. With the damping loss of the metal, $\epsilon_{\scriptscriptstyle \bot}^{\rm loc}$ and $\epsilon_{\scriptscriptstyle \bot}^{\rm nloc}$ show large but finite values near $\omega_{\rm res}^{\rm loc}$ and $\omega_{\rm res}^{\rm nloc}$, respectively. Around $\omega_{\rm res}^{\rm loc}$ and $\omega_{\rm res}^{\rm nloc}$, $\epsilon_{\scriptscriptstyle \bot}^{\rm loc}$ and $\epsilon_{\scriptscriptstyle \bot}^{\rm nloc}$ show a huge difference, indicating the importance of nonlocal response, consistent with the theoretical prediction in the previous section. Far from $\omega_{\rm res}^{\rm loc}$ and $\omega_{\rm res}^{\rm nloc}$, the dielectric functions $\epsilon_{\scriptscriptstyle \bot}^{\rm loc}$ and $\epsilon_{\scriptscriptstyle \bot}^{\rm nloc}$ are nearly equal, so that away from these resonances the nonlocal response is only a small perturbation.

\begin{figure}[h]
\centering
\includegraphics[width=0.6\textwidth]{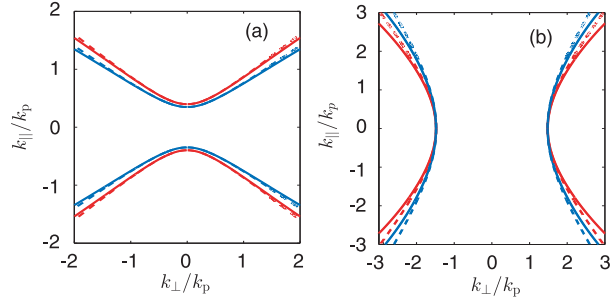}
\caption
{Dispersion curves of the HMM at (a) $\omega=0.1\omega_p$ and (b) $\omega=0.6\omega_p$. The HMM is as that in Fig.~\ref{fig2} except that the loss is neglected. Red solid curves for the exact local dispersion, blue solid curves for the exact nonlocal dispersion, red dashed curves for the approximated local dispersion of Eq.~(\ref{approx_dispersion2}), and blue dashed curves for the approximated nonlocal dispersion of Eq.~(\ref{approx_dispersion3}).}
\label{fig3}
\end{figure}

Fig.~\ref{fig3}(a) and (b) illustrate the dispersion curves  at $\omega=0.1\omega_{\rm p}$ and $\omega=0.6\omega_{\rm p}$, respectively. These frequencies are far away from the resonances $\omega_{\rm res}^{\rm loc}$ and $\omega_{\rm res}^{\rm nloc}$. The small damping loss is neglected for a clearer illustration of the dispersion curve. Several observations can be extracted from Fig.~\ref{fig3}. Firstly, the exact local and nonlocal dispersion curves agree well with those based on the effective material parameters, which confirms the validity of the approximate theoretical results of  Sec.~\ref{Sec:effective}. Secondly, the local and nonlocal curves only show a slight difference, agreeing well with Fig.~\ref{fig2}(c) where we see $\epsilon_{\scriptscriptstyle \bot}^{\rm nloc}\approx\epsilon_{\scriptscriptstyle \bot}^{\rm loc}$ both for $\omega=0.1\omega_{\rm p}$ and $0.6\omega_{\rm p}$.
\begin{figure}[h]
\centering
\includegraphics[width=0.6\textwidth]{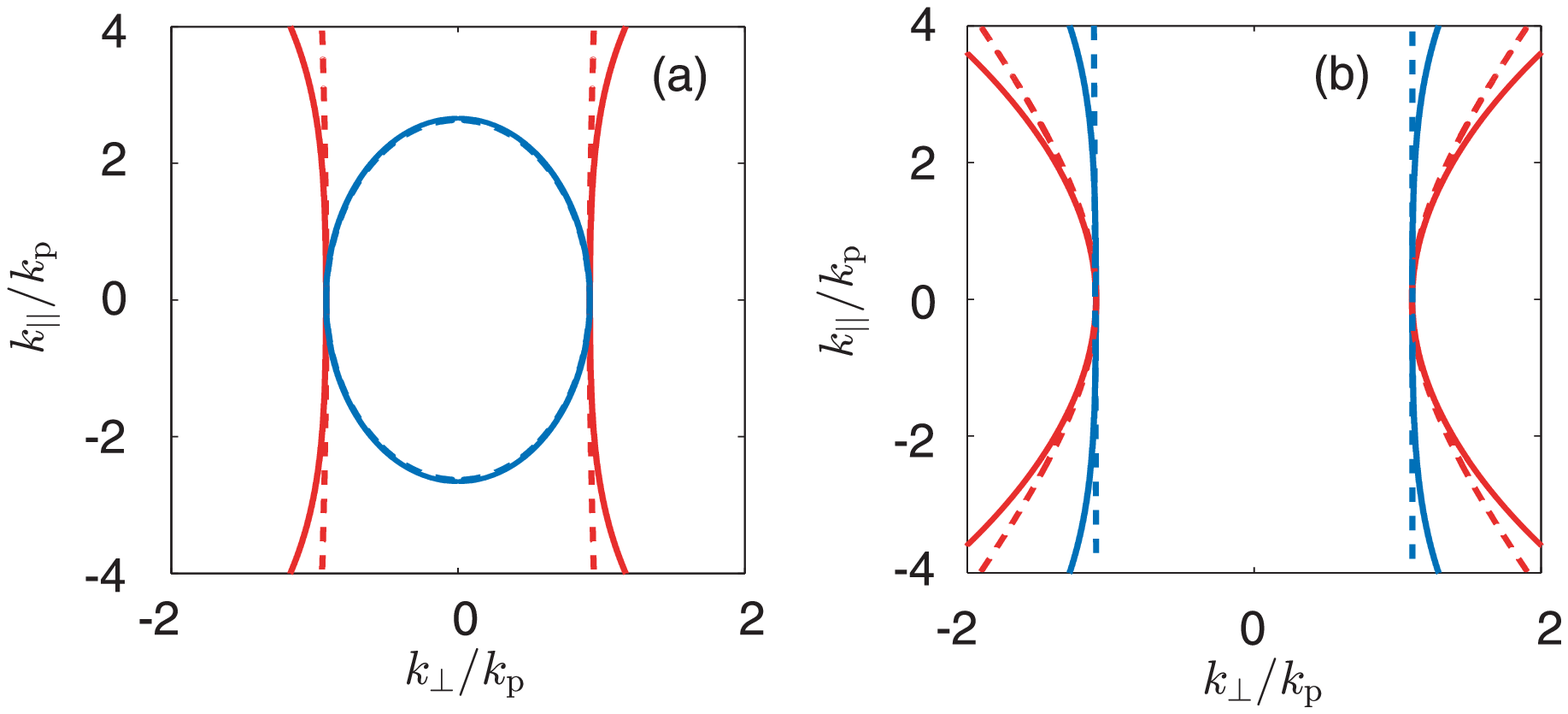}
\caption
{Dispersion curves of the HMM at (a) $\omega=\omega_{\rm res}^{\rm loc}=0.41\omega_p$ and (b) $\omega=\omega_{\rm res}^{\rm nloc}=0.47\omega_p$. The HMM is as that in Fig.~\ref{fig2} except that the loss is neglected. Red solid curves denote the exact local dispersion, blue solid curves the exact nonlocal dispersion; red dashed curves for the approximated local dispersion curves of Eq.~(\ref{approx_dispersion2}), and blue dashed curves for the approximated dispersion curves of Eq.~(\ref{approx_dispersion3}).}
\label{fig4}
\end{figure}

Fig.~\ref{fig4}(a) and (b) show the on-resonance dispersion curves, namely for $\omega=\omega_{\rm res}^{\rm loc}=0.41\omega_p=$ and for $\omega=\omega_{\rm res}^{\rm nloc}=0.47\omega_p$, respectively. Loss is again neglected,  as in Fig.~\ref{fig3}. At both frequencies, the nonlocal response modifies the dispersion curve noticeably. In particular, at $\omega=0.41\omega_{\rm p}$, the local dispersion curve consists of two nearly flat lines of $k_{\scriptscriptstyle \bot}=\pm k_{0}\sqrt{\epsilon_{\scriptscriptstyle \parallel}^{\rm loc}}$. With the nonlocal response, however, the dispersion curve becomes a closed ellipse! This remarkable difference can be understood from the difference between $\epsilon_{\scriptscriptstyle \bot}^{\rm loc}$ and $\epsilon_{\scriptscriptstyle \bot}^{\rm nloc}$. In particular, $\epsilon_{\scriptscriptstyle \bot}^{\rm loc}$ diverges while $\epsilon_{\scriptscriptstyle \bot}^{\rm nloc}=41$ stays finite in the lossless case. When including the loss, the effective parameters change to $\epsilon_{\scriptscriptstyle \bot}^{\rm loc}=-200 + 500i$ and $\epsilon_{\scriptscriptstyle \bot}^{\rm nloc}= 41+ 2.5i$. The loss would only slightly modify the dispersion curves of Fig.~\ref{fig4}(a). In Fig.~\ref{fig4}(b) we display dispersion relations at $\omega=0.47\omega_{\rm p}$. Here the local dispersion curve is a hyperbola. With the nonlocal response, however, the dispersion curve becomes nearly flat lines, which can be attributed to the extremely large value of $\epsilon_{\scriptscriptstyle \bot}^{\rm nloc}$ at this frequency $\omega_{\rm res}^{\rm nloc}$.

\section{Effects of nonlocal response on a hyperbolic metamaterial lens}
A HMM slab can operate as a superlens with subwavelength resolution, since the hyperbolic dispersion curve supports propagating waves with arbitrarily high wavevectors, which can transfer the evanescent information of the object. In the LRA, it is known that the HMM with its flat dispersion curve at $\omega_{\rm res}^{\rm loc}$ [recall Fig.~\ref{fig4}(a)] is especially
favorable for subwavelength imaging~\cite{Belov:2006}. The reason is that the flat dispersion curve with the constant $k_{\scriptscriptstyle\bot}$ ensures that all plane-wave components experience the same phase changes after transmission through the HMM slab, at least if reflections can be neglected. Actually,
the reflections can be suppressed by appropriate choice of the thickness $l$ of the HMM  lens, and even be made to vanish by choosing  $k_{\scriptscriptstyle \bot}l=n\pi$, where $n$ is a positive integer.
In theory this could lead to a perfect image at $x=0$.

Let us now investigate how nonlocal response may influence the subwavelength imaging characteristics of the HMM lens. As demonstrated in the above sections, the nonlocal response sets the infinite $\epsilon_{\scriptscriptstyle \bot}^{\rm loc}\to\infty$ to a finite value and accordingly destroys the desired flat dispersion curve at the frequency $\omega_{\rm res}^{\rm loc}$ where one would normally choose to operate for perfect imaging, based on the LRA. Nevertheless, at a blueshifted frequency $\omega_{\rm res}^{\rm nloc}$, the local effective dielectric function $\epsilon_{\scriptscriptstyle \bot}^{\rm loc}$ may be finite, but the hydrodynamic Drude model predicts instead that $\epsilon_{\scriptscriptstyle \bot}^{\rm nloc}$ diverges, with the concomitant flat dispersion curve suitable for subwavelength imaging. Thus nonlocal response is expected to strongly affect the performance of HMM superlenses for frequencies $\omega$ around $\omega_{\rm res}^{\rm loc}$ and $\omega_{\rm res}^{\rm nloc}$.

\begin{figure}[ht]
\centering
\includegraphics[width=0.6\textwidth]{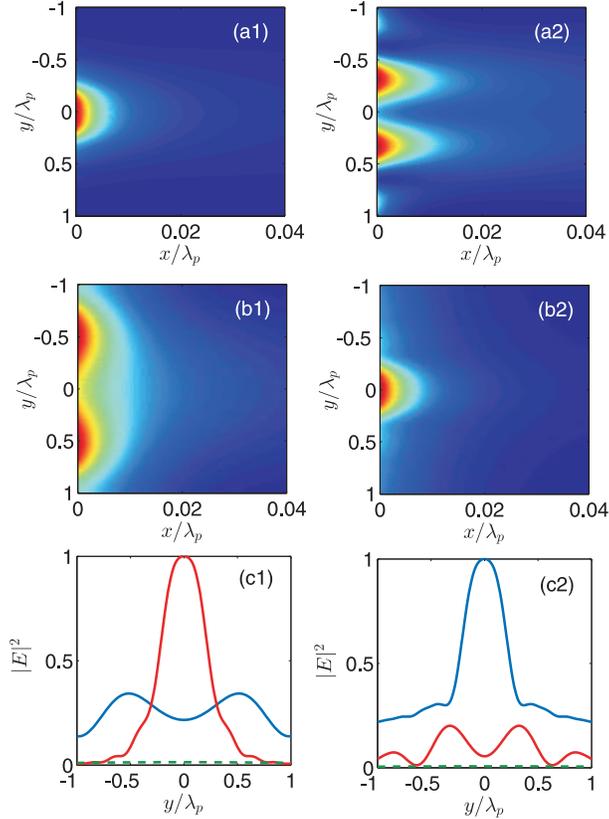}
\caption
{Transmitted electric-field intensity distribution  for a line dipole source $\mathbf J=\delta(x+l+x_{\rm s})\delta(y)\hat y$ positioned to the left of a HMM slab that has its left interface at $x=-l$ and right interface at $x=0$. For $\omega=0.41\omega_{\rm p}\approx\omega_{\rm res}^{\rm loc}$, panel (a1) shows the intensity for local response, (b1) for nonlocal response, and (c1) the intensity along $y$ with $x=0$.
Panels (a2), (b2), and (c2) are the analogous graphs for $\omega=0.47\omega_{\rm p}\approx\omega_{\rm res}^{\rm nloc}$.
In panels (c1) and (c2), the green dashed curves are for the case without the HMM slab, the red solid curves for the  HMM with local response, and the blue solid curves for the  HMM with nonlocal response. The HMM unit cell is as that in Fig.~\ref{fig2}, $x_{\rm s}=1\rm nm$ and $l=36d$.}
\label{fig5}
\end{figure}

As an example, we consider a HMM slab with $l=36d$, {\em i.e.} composed of $36$ unit cells, in a free-space background with the two boundaries at $x=-l$ and $x=0$.
The unit cell is as in Fig.~\ref{fig2}, and is arranged in a symmetric sandwich structure with the metal layer at the center. A line dipole source is positioned to the left of the HMM slab, with $x$-coordinate $-l-x_{\rm s}$ and $y$-coordinate 0, and is represented by $\mathbf J=\delta(x+l+x_{\rm s})\delta(y)\hat y$. We choose the distance to the HMM slab to be $x_{\rm s}=1\rm nm$.

Fig.~\ref{fig5}(a1) and (b1) demonstrate the transmitted electric-field intensity at $\omega=\omega_{\rm res}^{\rm loc}=0.41\omega_{\rm p}$ for local and nonlocal response, respectively. Metal loss is taken into account. In the local case, we have the known the dispersion of Fig.~\ref{fig4}(a) featuring flat lines. Additionally, the reflection is nearly zero since $k_{\scriptscriptstyle \bot}l\approx4\pi$, so all wave components experience the same phase change. Accordingly, after propagation through the HMM from $x=-l$ to $x=0$, a subwavelength image is formed near $x=0$, as seen in Fig.~\ref{fig5}(a1). By contrast, in the nonlocal case the dispersion curve turns into an ellipse. This closed dispersion curve sets an upper wavevector cutoff for the evanescent waves, and the wave components below the cutoff experience different phase changes. Accordingly, the quality of the image becomes worse, see Fig.~\ref{fig5}(a2).

In Fig.~\ref{fig5}(c1), we depict the electric-field intensity again at $\omega_{\rm res}^{\rm loc}$ as a function of $y$ at the HMM boundary $x=0$, and also for the case without the HMM slab. It is seen that the electric-field intensity is nearly vanishing in the absence of the HMM slab, since the evanescent components vanish after traveling a distance of $l$, which is of the order of one free-space wavelength. With the HMM slab in place, the electric-field intensity in the local case shows a subwavelength image that peaks at $y=0$, with a full width at half maximum (FWHM) of only  $0.47\lambda_{\rm p}$. With nonlocal response, however, the electric-field intensity distribution for $\omega_{\rm res}^{\rm loc}$ becomes flatter, with a double rather than a single peak, with peak intensities at $y=\pm0.55\lambda_{\rm p}$.

Let us now turn to the other resonance frequency, namely $\omega=\omega_{\rm res}^{\rm nloc}$ of Eq.~(\ref{res2}). Fig.~\ref{fig5}(a2) and (b2) show the transmitted electric-field intensity for the local and nonlocal cases, respectively, at $\omega=\omega_{\rm res}^{\rm nloc}=0.47\omega_p$. At this frequency, the local dispersion curve is a hyperbola, while the nonlocal dispersion shows two flat lines. In the nonlocal case, we have $k_{\scriptscriptstyle \bot}l\approx4.7\pi$ indicating that reflections do exist and the wave components experience different phase changes. However, compared to the local case, the transmission coefficients for the different wave components in the nonlocal case vary more smoothly as a function of ${\bf k}_{\parallel}$, thanks to the flat dispersion curve with the constant $k_{\scriptscriptstyle\bot}$.
As a result, the hydrodynamic Drude model predicts a better focusing performance of the HMM slab at $\omega=0.47\omega_p=\omega_{\rm res}^{\rm nloc}$ than does the local-response theory, compare Fig.~\ref{fig5}(a2) and (b2).
In Fig. \ref{fig5}(c2), the electric-field intensity along $y$ at $x=0$ is shown. In the local case, the electric-field intensity shows several peaks
with the strongest two at $y=\pm 0.33\lambda_p$. In the nonlocal case, the electric-field intensity is peaked at $y=0$ with a subwavelength FWHM of only $0.48\lambda_{\rm p}$.

\section{Detecting nonlocal response by near-field measurement}
\begin{figure}[ht]
\centering
\includegraphics[width=0.6\textwidth]{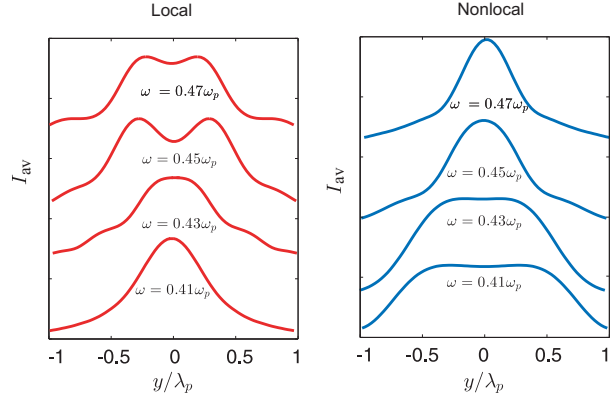}
\caption
{Calculated detector-area-averaged near-field intensity $I_{\rm av}$ of the light emitted by the same source as in Fig.~\ref{fig5} and transmitted through the same HMM slab. The detector is square-shaped with size $40\rm nm\times40\rm nm$, touching the HMM right interface at $x=0$. It scans along the $y$-direction, and the $y$-coordinate on the horizontal axes in panels~(a) for the LRA and (b) for the HDM corresponds to that of the center of the detector.
\label{fig6}}
\end{figure}

Fig.~\ref{fig5} indicates the possibility of detecting the nonlocal response experimentally by measuring the transmitted near-field distribution at the surface of a HMM superlens, which in our setup would be its right interface at $x=0$. The images in Fig.~\ref{fig5}(c1,c2) correspond to hypothetical measurements  with infinitely small detectors. In experiments, the measured near-field signal will  rather be an area-averaged electric-field intensity
\begin{equation}
I_{\rm av} \propto \frac{1}{D}\int\limits_D \mbox{d}^{2}{\bf r}\,|{\mathbf E}({\bf r})|^2,
\end{equation}
where $D$ is the finite detection area of the detector. Let us now assume that we have a near-field detector with detection area in the $yz$-plane, with a square shape of size $40\,\rm nm\times40\,\rm nm$, touching the HMM interface at $x=0$.  For the same light source interacting with the HMM slab as in Fig.~\ref{fig5}, we depict in Fig.~\ref{fig6} the calculated $I_{\rm av}$ as a function of the $y$-coordinate of the center of the detector. It is seen that local and nonlocal response models give significantly different predictions for the measured signal, and thus the differences between the two models survive detection area averaging. The single peaks at $0.41\omega_{\rm p}$ for local response and at $0.47\omega_{\rm p}$ for nonlocal response are naturally broader than in Fig.~\ref{fig5}(c1,c2), also due to the area averaging. Interestingly, as the frequency increases from $0.41\omega_{\rm p}$ to $0.47\omega_{\rm p}$, the distribution of $I_{\rm av}$ becomes broader in the LRA, but narrower in the hydrodynamic Drude model.

\section{Conclusions}
We investigated the effects of the hydrodynamic nonlocal response on hyperbolic metamaterials with  periodicity in the subwavelength regime of the transverse optical wave, but much larger than the wavelength of the longitudinal hydrodynamic pressure waves. It is found that the nonlocal response corrects the effective permittivity tensor element $\epsilon_{\scriptscriptstyle \bot}^{\rm loc}$ in the periodicity direction to the new form $\epsilon_{\scriptscriptstyle \bot}^{\rm nloc}$ of Eq.~(\ref{nlocapmp}). Around the frequencies $\omega_{\rm res}^{\rm loc}$ and $\omega_{\rm res}^{\rm nloc}$ corresponding to $\epsilon_{\scriptscriptstyle \bot}^{\rm loc}\to\infty$ and $\epsilon_{\scriptscriptstyle \bot}^{\rm nloc}\to\infty$, respectively, $\epsilon_{\scriptscriptstyle \bot}^{\rm loc}$ and $\epsilon_{\scriptscriptstyle \bot}^{\rm nloc}$ show noticeable differences,
even leading to completely different dispersion curves with and without the nonlocal response.

We find that nonlocal response blueshifts the resonance frequency of HMMs from $\omega_{\rm res}^{\rm loc}$ [Eq.~(\ref{res1})] to $\omega_{\rm res}^{\rm nloc}$ [Eq.~(\ref{res2})]. The relative blueshift has an interesting simple form independent of the thickness of the metal layers. Similar nonlocal blueshifts for single nanoplasmonic particles have been predicted before, and significant blueshifts have also been measured~\cite{Scholl:2012,Raza:2012b}; how much of these can be attributed to hydrodynamic effects is a hot topic~\cite{Scholl:2012,Raza:2012b,Teperik:2013}.

Furthermore, we predict that nonlocal response shows its mark in the performance of a finite HMM slab as  a focusing lens: when increasing the operating frequency from $\omega_{\rm res}^{\rm loc}$ to $\omega_{\rm res}^{\rm nloc}$, for local response the near-field distribution of the transmitted light shows an image that gets out of focus, whereas the focus would improve instead according to the nonlocal-response theory. We propose to test the blueshift and these contrary predictions experimentally, as a clear and interesting test whether nonlocal response can be observed in hyperbolic metamaterials.

\section*{Acknowledgments}
This work was financially supported by an H.~C. {\O}rsted Fellowship (W.Y.).
\end{document}